\documentclass[twocolumn,showpacs,prl]{revtex4}
\usepackage{times}
\usepackage{graphicx}
\usepackage{graphics}
\usepackage{dcolumn}
\usepackage{amsmath}
\def\lambdabar{\protect\@lambdabar}
\def\@lambdabar{%
\relax \bgroup
\def\@tempa{\hbox{\raise.73\ht0
\hbox to0pt{\kern.25\wd0\vrule width.5\wd0
height.1pt depth.1pt\hss}\box0}}%
\mathchoice{\setbox0\hbox{$\displaystyle\lambda$}\@tempa}%
{\setbox0\hbox{$\textstyle\lambda$}\@tempa}%
{\setbox0\hbox{$\scriptstyle\lambda$}\@tempa}%
{\setbox0\hbox{$\scriptscriptstyle\lambda$}\@tempa}%
\egroup }
\begin{document}
\title{Photon Recoil Momentum in Dispersive Media}

\author{Gretchen K. Campbell, Aaron E. Leanhardt\footnote{Present address: JILA, Boulder, CO 80309. }, Jongchul Mun, Micah Boyd, Erik W. Streed, Wolfgang Ketterle and David E.
Pritchard} \homepage[URL: ]{http://cua.mit.edu/ketterle_group/}
\affiliation{MIT-Harvard Center for Ultracold Atoms, Research
Laboratory of Electronics and Department of Physics, Massachusetts
Institute of Technology, Cambridge, MA 02139, USA}
\date{\today}

\begin{abstract}
A systematic shift of the photon recoil due to the index of
refraction of a dilute gas of atoms has been observed. The recoil
frequency was determined with a two-pulse light grating
interferometer using near-resonant laser light. The results show that
the recoil momentum of atoms caused by the absorption of a photon is
$n\hbar k$, where $n$ is the index of refraction of the gas and $k$
is the vacuum wavevector of the photon. This systematic effect must
be accounted for in high-precision atom interferometry with light
gratings.

\end{abstract}

\maketitle

The momentum of a photon in a dispersive medium is of conceptional
and practical importance
\cite{Minkowski08,Minkowski10,Abraham09,Abraham10,Gordon73,Haugan82}.
When a photon enters a medium with index of refraction $n$, the
electromagnetic momentum changes from $\hbar k$ to $\hbar k/n$,
where k is the vacuum wavevector of the photon. Momentum
conservation requires that the medium has now a mechanical momentum
$(1-1/n)\hbar k$ which for $n>1$, is parallel to the propagation of
the light. The particles in the medium are accelerated by the
leading edge of the light pulse and decelerated by the trailing
edge. As a result, no motion is left in the medium after the pulse
has passed. An absorbing surface without reflection would therefore
receive a momentum of $\hbar k$ per incident photon. In contrast, a
reflecting surface will recoil with a momentum of $2 n \hbar k$ per
photon. In this case, the standing wave formed by the incident and
reflected light pulse transfers momentum to the medium which remains
even after the light pulse has left. This modification of the recoil
momentum has so far only been observed for light being reflected
from a mirror immersed in a liquid \cite{Jones54,Jones78}.

Recently, there have been discussions about what happens to an atom
when it absorbs a photon within the medium. If one assumes that
after absorbing the photon, no motion is left in the medium, then
the recoil momentum should be $\hbar k$ \cite{Hensley01}. The same
conclusion is reached when one assumes a very dilute, dispersive
medium with the absorbing atom localized in the vacuum space between
the particles of the medium \cite{Ketterle}. However, the correct
answer is that the atom will recoil with a momentum of $n \hbar k$,
which requires particles in the medium to receive a backward
momentum (for $n>1$) due to the interaction of the oscillating
dipole moments of the particles in the dispersive medium and the
absorbing atom. So both for reflection and absorption by an atom, a
photon in a dispersive medium behaves as if it has a momentum of $n
\hbar k$.

In this paper, we study the recoil momentum of a photon in a dilute
gas and show that its value is $n \hbar k$.  This has important
consequences for atom interferometers using optical waves to
manipulate atoms by the transfer of recoil momentum. High precision
measurements of the photon recoil are used to determine the fine
structure constant $\alpha$
\cite{Taylor94,Weiss93,Wicht02,Gupta02,Battesti04,Coq05}. Further
improvements in the accuracy of photon recoil measurements, combined
with the value of $\alpha$ derived from the (g-2) measurements for
the electron and positron \cite{VanDyck87,Kinoshita95,Hughes99},
would provide a fundamental test of QED. The accuracy of the best
photon recoil measurements are limited by the uncertainty in the
correction to the photon recoil due to the index of refraction. At
low atomic densities, where atom interferometers usually operate,
the index of refraction effect is relatively small.  Here we operate
an atom interferometer with Bose-Einstein condensates, which have
much higher density than laser cooled atomic clouds, and observe how
the index of refraction modifies the atomic recoil frequency
$\hbar\omega_{rec}=\frac{p^2}{2m}$, where $p$ and $m$ are the atomic
recoil momentum and mass respectively.

\begin{figure}
\centering{
\includegraphics[hiresbb=true]{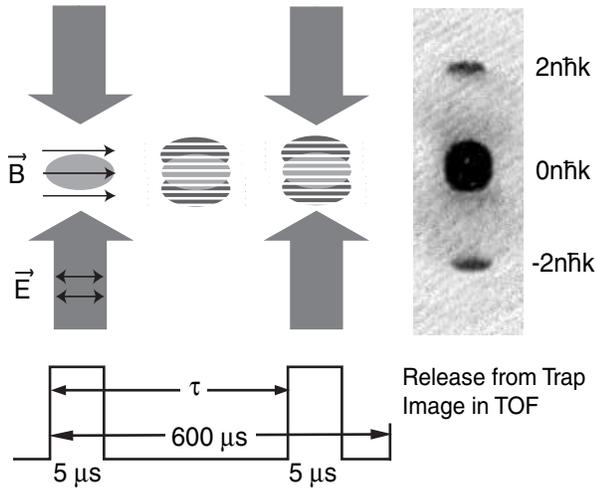}
\caption{\label{fig:figthree} Kapitza-Dirac interferometer. The
first pulse outcoupled a small fraction of atoms into the
$|\pm2n\hbar k\rangle$ momentum states. The outcoupled atoms moved
within the initial condensate. After a variable delay $\tau$ a
second pulse was applied and outcoupled atoms that interfered with
those outcoupled by the first pulse. The laser beam was applied
perpendicular to the long axis of the condensate; the polarization,
$\vec{E}$, was parallel to it and to the applied magnetic field
bias, $\vec{B}$. 600 $\mu s$ after the first pulse was applied the
atoms were released from the magnetic trap and imaged after 38 ms of
ballistic expansion. The field of view is 0.5 mm $\times$ 1.5 mm}}
\end{figure}

The essential idea of our experiment is to measure the recoil
frequency interferometrically using a two-pulse Ramsey
interferometer.  The two pulses are optical standing waves separated
by a delay time $\tau$ (Fig.~\ref{fig:figthree}). The first pulse
diffracts the atoms in an $^{87}$Rb condensate into discrete
momentum states. During the delay time $\tau$ the phase of each
momentum state evolves at a different rate according to its recoil
energy. The second pulse recombines the atoms with the initial
condensate. The recombined components have differing phases leading
to interference fringes that oscillate at the two-photon recoil
frequency. By measuring the resulting frequency as a function of the
standing wave detuning from the atomic resonance, we found a
distinctive dispersive shape for $\omega_{rec}$ that fit the recoil
momentum as $n\hbar k$, where $n$ is the index of refraction of the
gas.

The experiment was performed using an elongated $^{87}$Rb
Bose-Einstein condensate created in a cloverleaf-type
Ioffe-Pritchard magnetic trap using the procedure described in
Ref.\cite{Schneble03}. The condensate, containing $1.5\times10^6$
atoms, was produced in the $|5^2S_{1/2},F=1, m_F=-1\rangle$ state,
and had a Thomas-Fermi radius of 8 (90) $~\mu$m in the radial
(axial) direction, and the magnetic trap had a radial (axial) trap
frequency of 81 (7) Hz.

The BEC was illuminated from the side with an optical standing wave
created by a retro-reflected, $\pi$-polarized laser beam. The
polarization of the beam was optimized by suppressing Rayleigh
superradiance \cite{Schneble03}. The laser was detuned from the
$5^2S_{1/2},F=1\to 5^2P_{3/2},F=1$ transition at $\lambda=780$~nm,
and had a linewidth $\gamma$ much smaller than $\Gamma$, the natural
linewidth of the transition. The intensity of the 5 $\mu$s long
pulse was set to outcouple $\approx 5\%$ of the atoms i
nto each of
the $|\pm2n\hbar k\rangle$ momentum states with no appreciable
population in higher momentum states. This ensured that the density
of the original condensate was nearly constant throughout the
measurement. After a variable time $\tau$, a second identical pulse
was applied. The time between the first pulse and the shutoff of the
magnetic trap was fixed at 600 $\mu$s, which was less than a quarter
of the radial trap period. The momentum distribution of the
condensate was imaged after 38 ms of ballistic expansion, long
enough for the momentum states to be resolved. The images were
obtained using resonant absorption imaging after first optically
pumping the atoms to the $5^2S_{1/2},F=2$ state. To compensate for
spontaneous light scattering from the standing wave, the density of
the condensate (and associated mean field shift) was determined by
applying a single 5 $\mu s$ pulse to the condensate, and then
immediately releasing it from the magnetic trap. The number of atoms
in the condensate was determined by integrating the optical density
of the absorption image. The optical density was calibrated by
fitting the Thomas-Fermi radius of unperturbed condensates in
time-of-flight \cite{Castin96}.

\begin{figure}
\centering{
\includegraphics[hiresbb=true]{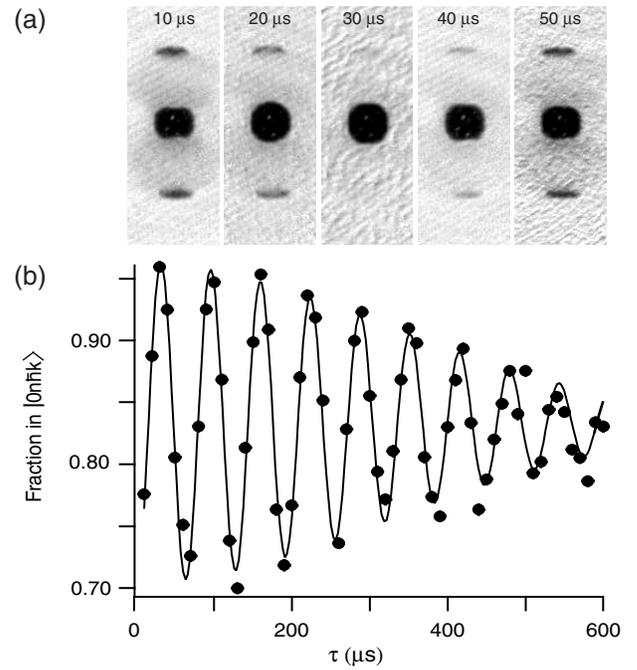}
\caption{\label{fig:figone} Interference fringes oscillating at the
recoil frequency. (a) Absorption images for $\tau=10-50$ $\mu$s. The
detuning was $\Delta/2\pi =+520$ MHz relative to the F=1 $\to$
F$'$=1 transition. The field of view is 0.5 mm $\times$ 1.5 mm (b)
Fraction of atoms in the $|0\hbar k\rangle$ momentum state as a
function of $\tau$. The fringes were fit using Eq.~(\ref{eq:six}).
The fitted frequency was $\omega=2\pi\times156268(39)$ Hz and the
decay constant $\tau_c=461(25)$ $\mu$s. The signal was normalized
using the total atom number in all momentum states.}}
\end{figure}

The recoil frequency was found by fitting the oscillations in the
fraction of atoms in the $|0n\hbar k\rangle$ momentum state for
$\tau$ ranging from $10 - 590$ $\mu s$ (Fig.~\ref{fig:figone}). The
interference fringes were fit to a cosine function with a gaussian
envelope:
\begin{equation}
\label{eq:six} A \exp\left({-\frac{\tau
^2}{\tau_c^2}}\right)\cos(4\omega\tau +\phi)+C.
\end{equation}
The observation of up to ten oscillations provided a precise value of
the recoil frequency.   The origin of the damping time $\tau_c$ and
of the offset $C$ will be discussed later.

\begin{figure}
\centering{
\includegraphics[hiresbb=true]{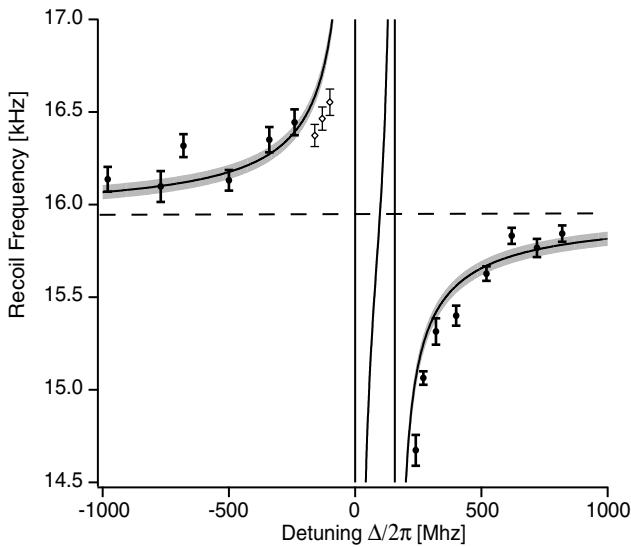}
\caption{\label{fig:figtwo} Recoil frequency as a function of
detuning $\Delta/2\pi$ showing the dispersive effect of the index of
refraction. The average density of the condensate for the solid
points was $1.14(4)\times 10^{14}$~cm$^{-3}$, giving rise to a
mean-field shift of 880 Hz. The shaded area gives the expected
recoil frequency including the uncertainty in the density. The
dashed line is at $\omega=4n^2\omega_{rec}+\rho U/\hbar$, the
expected value without index of refraction effects. The data shown
as open diamonds had increased spontaneous light scattering due to
$\sigma^{\pm}$ light contamination in the laser beam. The increased
light scattering led to a lower initial density in the condensate,
thus leading to a smaller mean-field shift and index of refraction.
The $\sigma^{\pm}$ contamination allowed $\Delta m_F=\pm1$
transitions, thus for small detunings the proximity to the
$|1,-1\rangle$ $\to$ $|0',0\rangle$ transition located at
$\Delta/2\pi=-72 $ MHz resulted in higher spontaneous scattering
rates. The open points have been scaled upward to correct for this
lower density.}}
\end{figure}

Figure~\ref{fig:figtwo} shows our measured values for $\omega/2\pi$
as a function of the detuning, $\Delta/2\pi$, relative to the F=1
$\to$ F$'$=1 transition. The measured values for the recoil
frequency clearly follow the dispersive shape of the index of
refraction, and they are in good agreement with the hypothesis that
the recoil momentum is proportional to the index of refraction. The
variation in $\omega$/$2\pi$ as a function of the detuning was 2 kHz
across the resonance, much larger than the statistical error on the
frequency fits of less than 100 Hz.  This conclusively shows that
the momentum transferred to the atom when a photon is absorbed is
$n\hbar k$.

We now discuss in more detail how the atoms interact with optical
standing waves.  For the short duration of the applied pulses (5
$\mu s$) we can assume that the atoms do not move during the pulse
and ignore the kinetic energy of the atoms (Raman-Nath
approximation). The interaction can then be described by the
application of the AC Stark potential due to the standing wave
\begin{equation}
V(z)=\frac{\hbar\omega_{R}^2}{\Delta}\sin^2(nkz),
\end{equation}
where $\Delta$ is the detuning between the optical frequency and the
atomic transition, and $\omega_{R}$ is the Rabi frequency. This
equation is valid for large detuning, $\Delta^2\gg\Gamma^2/4$, where
$\Gamma$ is the natural linewidth of the transition. The short pulse
limit, describing Kapitza-Dirac scattering, is valid for short
interaction times $t_p$, where $t_p\ll1/\omega_{rec}\approx40\mu$s.
The first pulse outcouples a fraction of atoms into the momentum
states $|\pm2\ell n\hbar k\rangle$, where the population in the
$\ell^{th}$ momentum state is given by $P_{\ell}=J^2_{\ell}(\theta)$
\cite{Meystre01,Gupta01}, where for a square pulse,
$\theta=\frac{\omega_{R}^2 t_p}{2\Delta}$, and $J_{\ell}$ is the
$\ell^{th}$-order Bessel Function of the first kind. For $\theta<1$
a negligible fraction of atoms is diffracted into states with
$\ell>1$, and we can restrict our discussion to the $|\pm2n\hbar
k\rangle$ momentum states. For our experimental parameters
$\theta=0.45$. During the delay time $\tau$ the phase of the
$|\pm2n\hbar k\rangle$ states evolves at a faster rate than the
$|0n\hbar k\rangle$ state due to the recoil energy,
$E_{rec}=4n^2\hbar\omega_{rec}$, hence the wavefunction will evolve
as
\begin{equation}
|\psi(\tau)\rangle=|\psi_o\rangle\left(J_1(\theta)|\pm2n\hbar
k\rangle e^{-i4n^2\omega_{rec}\tau}+J_0(\theta)|0n\hbar
k\rangle\right).
\end{equation}
At $t=\tau$ a second pulse is applied that partially recombines the
momentum states. After applying the two pulses, the probability of
finding the atoms in the the $|0n\hbar k\rangle$ momentum state,
$\rho_0=|\langle\psi(\tau+t_p)|0n\hbar k\rangle|^2$ is given by
\begin{equation}
\rho_0=J_0^4(\theta)+4\left[J_0^2(\theta)J_1^2(\theta)+J_1^4(\theta)\right]\cos(4n^2\omega_{rec}\tau).
\end{equation}
As a function of $\tau$ the density of the zero momentum peak
oscillates at $4n^2\omega_{rec}$.

So far, we have ignored the motion of the atoms during the delay time
$\tau$.  However the amplitudes of the recombined components will
only interfere where they spatially overlap. After the first pulse,
the atoms in the $|\pm2n\hbar k\rangle$ momentum states move within
the initial condensate with the recoil velocity ($\sim12$ $\mu$m/ms).
As the overlap between the recoiling atoms and those at rest
decreases, the interference fringes decay. The overlap integral for
this decay can be approximated as a Gaussian with time constant,
$\tau_c\approx$ 0.75 $R_{TF}/v_{rec}$, where $R_{TF}$ is the
Thomas-Fermi radius of the condensate, and $v_{rec}$ is the recoil
velocity \cite{Trippenbach00}.

The index of refraction for the condensate is derived from its
macroscopic polarization $P$. For a two level system,
\begin{equation}
P=\chi\epsilon_oE=i\rho\frac{\mu^2}{\hbar}\frac{E}{\Gamma-i\Delta}
\end{equation}
where $\chi$ is the atomic susceptibility, $\epsilon_o$ is the
permittivity of free space, $\mu$ is the dipole matrix element,
$\rho$ is the atomic density of the condensate. In this experiment
the light was $\pi$-polarized and detuned by $\Delta$ from the
$5^2S_{1/2},F=1\to5^2P_{3/2},F'=1$ transition. For this polarization
the selection rule is $\Delta m_F=0$, and there are two allowed
transitions from $|F=1,m_F=-1\rangle$ $\to$ $|1',-1\rangle$ and
$|1,-1\rangle$ $\to$ $|2',-1\rangle$ which are separated by 157 MHz.
Including both transitions in the derivation, the index of
refraction, $n=\sqrt{1+\chi}$, is given by:
\\
\\ $n$=
\begin{equation}
\sqrt{1-12\pi\rho\left(\frac{\lambda}{2\pi}\right)^3\left(\frac{5}{12}\frac{\Delta_1}{\Gamma}\frac{1}{1+(2
\frac{\Delta_1}{\Gamma})^2}+\frac{1}{4}\frac{\Delta_2}{\Gamma}\frac{1}{1+(2\frac{\Delta_2}{\Gamma})^2}\right)}
\end{equation}
where $\Delta_1$ and $\Delta_2$ are the detunings relative to the
$F=1 \to F'=1$ and $F'=2$ transitions, respectively. This equation
is valid in the limit $\rho\lambdabar^3\ll1$ \cite{Morice95}, where
$\lambdabar=\lambda/2\pi$. For our experimental parameters
$\rho\lambdabar^3\approx0.2$. In addition to the index of refraction
shift, the observed recoil frequency has a mean field shift
\cite{Stenger99}; the atoms in the $|\pm2n\hbar k\rangle$ state have
twice the mean field energy of those at rest due to the exchange
term in the interatomic potential. Including both the mean field
shift and the index of refraction, the frequency of the observed
interference fringes should be:
\begin{equation}
\label{eq:seven} \omega=4n^2\omega_{rec}+\frac{\rho U}{\hbar}
\end{equation}
where $\rho U=4\pi\hbar^2a\rho/m$, and $a$ is the s-wave scattering
length. The density $\rho=4/7\rho_o$, where $\rho_o$ is the peak
condensate density and the factor of 4/7 is due to the inhomogeneous
condensate density.

When the interference fringes were fit using Eq.~(\ref{eq:six}), the
average values for the amplitude $A$ and offset $C$ for all of the
data points were 0.12(3) and 0.82(4), respectively. This is in
reasonable agreement with the expected values of $A$ = 0.18 and $C$
= 0.81 for $\theta=0.45$. For a Thomas-Fermi radius of 8 $\mu$m we
would expect a decay time $\tau_c\approx$ $500$ $\mu$s. There was an
unexplained shift in the fitted value for the decay time $\tau_c$
between the red and blue side of the resonances, on the red side the
average value was 347(20) $\mu$s and on the blue, 455(40) $\mu$s.

The quadratic dependence in Eq.~(\ref{eq:seven}) on the index of
refraction can be understood by considering the diffraction of atoms
from the light grating created by the standing wave. When the first
pulse is applied, the standing wave creates a grating with a
periodicity $d=\lambda$/2. Within the condensate the index of
refraction modifies the grating period by $n$, since
$\lambda'=\lambda/n$. The recoil momentum of atoms that diffract off
the grating will be changed by $\pm 2\hbar k'$, again within the
condensate $k'=nk$ and the velocity of the atoms is modified.
Assuming $n>1$, when the second grating is applied the atoms have
moved farther by a factor of $n$ and the grating is shorter by $n$,
changing the time scale for the interference fringes by a factor of
$n^2$.

The increase in the momentum transferred to the atoms can also be
explained by considering the momentum transferred to atoms by a
classical field. A derivation using the Lorentz force applied to the
atoms during the absorption of a photon can be found in
\cite{Haugan82}.  In a dielectric medium with $n>1$, the magnetic
field and therefore the Lorentz force are not modified.  However, the
electric field is weaker and therefore it takes longer for the atom
to perform half a Rabi cycle and be transferred to the excited state.
During that longer time, the Lorentz force imparts a momentum to the
atom which is larger than $\hbar k$.

For Kapitza-Dirac scattering, atoms are diffracted symmetrically
into the into $|\pm2\ell n\hbar k\rangle$ momentum states, so
momentum is clearly conserved. However for processes such as Bragg
scattering, where the atoms are scattered in only one direction the
index of refraction has an additional effect. Assuming a $\pi/2$
pulse with counter-propagating beams, where half the atoms are
diffracted, for $n>1$ the recoil momentum is a factor of $n$ higher
than the vacuum momentum. For momentum to be conserved, the
remaining atoms must recoil backwards with momentum $p=2(n-1)\ell
\hbar k$. For small fractional outcoupling the effect is negligible,
since the extra momentum is distributed among the remaining
condensate. However if a large fraction of the condensate is
outcoupled and $\ell$ is large, this effect could potentially be
resolved in ballistic expansion.

We have discussed here the dispersive effect on the photon momentum
near a one-photon resonance.  An analogous effect occurs near
two-photon resonances. In this case, the atomic polarizability is
determined in third-order perturbation theory, and the resulting
index of refraction has a sharp, narrow dispersive feature near the
two-photon resonance \cite{Schirotzek04}. In recent experiments at
Stanford \cite{Wicht02}, such two-photon effects have been the
leading source of uncertainty in high-precision determinations of
atomic recoil frequencies and the fine-structure constant $\alpha$.

In conclusion we have measured a systematic shift in the photon
recoil frequency due to the index of refraction of the condensate.
This is the first direct observation of the atomic recoil momentum
in dispersive media. For high atomic densities, this shift can have
a significant effect on atom interferometers, and is of particular
importance for precision measurements of $h/m$ and $\alpha$ with
cold atoms \cite{Wicht02,Coq05}.

The authors thank A. Schirotzek, S. Chu, S. Harris and H. Haus for
insightful discussions, and M. Kellogg for experimental assistance.
This work was supported by NSF and ARO.


\begin{thebibliography}{27}
\expandafter\ifx\csname
natexlab\endcsname\relax\def\natexlab#1{#1}\fi
\expandafter\ifx\csname bibnamefont\endcsname\relax
  \def\bibnamefont#1{#1}\fi
\expandafter\ifx\csname bibfnamefont\endcsname\relax
  \def\bibfnamefont#1{#1}\fi
\expandafter\ifx\csname citenamefont\endcsname\relax
  \def\citenamefont#1{#1}\fi
\expandafter\ifx\csname url\endcsname\relax
  \def\url#1{\texttt{#1}}\fi
\expandafter\ifx\csname urlprefix\endcsname\relax\def\urlprefix{URL
}\fi \providecommand{\bibinfo}[2]{#2}
\providecommand{\eprint}[2][]{\url{#2}}

\bibitem[{\citenamefont{Minkowski}(1908)}]{Minkowski08}
\bibinfo{author}{\bibfnamefont{H.}~\bibnamefont{Minkowski}},
  \bibinfo{journal}{Nachr. Ges. Wiss. G\"{o}ttingen} p.~\bibinfo{pages}{53}
  (\bibinfo{year}{1908}).

\bibitem[{\citenamefont{Minkowski}(1910)}]{Minkowski10}
\bibinfo{author}{\bibfnamefont{H.}~\bibnamefont{Minkowski}},
  \bibinfo{journal}{Math. Ann} \textbf{\bibinfo{volume}{68}},
  \bibinfo{pages}{472} (\bibinfo{year}{1910}).

\bibitem[{\citenamefont{Abraham}(1909)}]{Abraham09}
\bibinfo{author}{\bibfnamefont{M.}~\bibnamefont{Abraham}},
  \bibinfo{journal}{Rend. Pal} \textbf{\bibinfo{volume}{28}},
  \bibinfo{pages}{1} (\bibinfo{year}{1909}).

\bibitem[{\citenamefont{Abraham}(1910)}]{Abraham10}
\bibinfo{author}{\bibfnamefont{M.}~\bibnamefont{Abraham}},
  \bibinfo{journal}{Rend. Pal} \textbf{\bibinfo{volume}{30}},
  \bibinfo{pages}{33} (\bibinfo{year}{1910}).

\bibitem[{\citenamefont{Gordon}(1973)}]{Gordon73}
\bibinfo{author}{\bibfnamefont{J.~P.} \bibnamefont{Gordon}},
  \bibinfo{journal}{Phys.\ Rev.\ A} \textbf{\bibinfo{volume}{8}},
  \bibinfo{pages}{14} (\bibinfo{year}{1973}).

\bibitem[{\citenamefont{Haugan and Kowalski}(1982)}]{Haugan82}
\bibinfo{author}{\bibfnamefont{M.~P.} \bibnamefont{Haugan}} \bibnamefont{and}
  \bibinfo{author}{\bibfnamefont{F.~V.} \bibnamefont{Kowalski}},
  \bibinfo{journal}{Phys.\ Rev.\ A} \textbf{\bibinfo{volume}{25}},
  \bibinfo{pages}{2102} (\bibinfo{year}{1982}).

\bibitem[{\citenamefont{Jones and Richards}(1978)}]{Jones54}
\bibinfo{author}{\bibfnamefont{R.~V.} \bibnamefont{Jones}} \bibnamefont{and}
  \bibinfo{author}{\bibfnamefont{J.~C.~S.} \bibnamefont{Richards}},
  \bibinfo{journal}{Proc. R. Soc. London} \textbf{\bibinfo{volume}{A221}},
  \bibinfo{pages}{480} (\bibinfo{year}{1978}).

\bibitem[{\citenamefont{Jones and Leslie}(1978)}]{Jones78}
\bibinfo{author}{\bibfnamefont{R.~V.} \bibnamefont{Jones}} \bibnamefont{and}
  \bibinfo{author}{\bibfnamefont{B.}~\bibnamefont{Leslie}},
  \bibinfo{journal}{Proc. R. Soc. London} \textbf{\bibinfo{volume}{A360}},
  \bibinfo{pages}{347} (\bibinfo{year}{1978}).

\bibitem[{\citenamefont{Hensley et~al.}(2001)\citenamefont{Hensley, Wicht,
  Young, and Chu}}]{Hensley01}
\bibinfo{author}{\bibfnamefont{J.~M.} \bibnamefont{Hensley}},
  \bibinfo{author}{\bibfnamefont{A.}~\bibnamefont{Wicht}},
  \bibinfo{author}{\bibfnamefont{B.~C.} \bibnamefont{Young}}, \bibnamefont{and}
  \bibinfo{author}{\bibfnamefont{S.}~\bibnamefont{Chu}}, in
  \emph{\bibinfo{booktitle}{Atomic Physics 17}}, edited by
  \bibinfo{editor}{\bibfnamefont{E.}~\bibnamefont{Arimondo}},
  \bibinfo{editor}{\bibfnamefont{P.~D.} \bibnamefont{Natale}},
  \bibnamefont{and} \bibinfo{editor}{\bibfnamefont{M.}~\bibnamefont{Inguscio}}
  (\bibinfo{year}{2001}).

\bibitem[{Ket()}]{Ketterle}
\bibinfo{note}{This argument has been made previously by one of the authors (W.
  K.)}.

\bibitem[{\citenamefont{Taylor}(1994)}]{Taylor94}
\bibinfo{author}{\bibfnamefont{B.}~\bibnamefont{Taylor}},
  \bibinfo{journal}{Metrologia} \textbf{\bibinfo{volume}{31}},
  \bibinfo{pages}{181} (\bibinfo{year}{1994}).

\bibitem[{\citenamefont{Weiss et~al.}(1993)\citenamefont{Weiss, Young, and
  Chu}}]{Weiss93}
\bibinfo{author}{\bibfnamefont{D.~S.} \bibnamefont{Weiss}},
  \bibinfo{author}{\bibfnamefont{B.~C.}~\bibnamefont{Young}}, \bibnamefont{and}
  \bibinfo{author}{\bibfnamefont{S.}~\bibnamefont{Chu}},
  \bibinfo{journal}{Phys.\ Rev.\ Lett} \textbf{\bibinfo{volume}{70}},
  \bibinfo{pages}{2706} (\bibinfo{year}{1993}).

\bibitem[{\citenamefont{Wicht et~al.}(2002)\citenamefont{Wicht, Hensley,
  Sarajlic, and Chu}}]{Wicht02}
\bibinfo{author}{\bibfnamefont{A.}~\bibnamefont{Wicht}},
  \bibinfo{author}{\bibfnamefont{J.~M.} \bibnamefont{Hensley}},
  \bibinfo{author}{\bibfnamefont{E.}~\bibnamefont{Sarajlic}}, \bibnamefont{and}
  \bibinfo{author}{\bibfnamefont{S.}~\bibnamefont{Chu}},
  \bibinfo{journal}{Physica Scripta} \textbf{\bibinfo{volume}{102}},
  \bibinfo{pages}{82} (\bibinfo{year}{2002}).

\bibitem[{\citenamefont{Gupta et~al.}(2002)\citenamefont{Gupta, Dieckmann,
  Hadzibabic, and Pritchard}}]{Gupta02}
\bibinfo{author}{\bibfnamefont{S.}~\bibnamefont{Gupta}},
  \bibinfo{author}{\bibfnamefont{K.}~\bibnamefont{Dieckmann}},
  \bibinfo{author}{\bibfnamefont{Z.}~\bibnamefont{Hadzibabic}},
  \bibnamefont{and} \bibinfo{author}{\bibfnamefont{D.~E.}
  \bibnamefont{Pritchard}}, \bibinfo{journal}{Phys.\ Rev.\ Lett}
  \textbf{\bibinfo{volume}{89}}, \bibinfo{pages}{140401}
  (\bibinfo{year}{2002}).

\bibitem[{\citenamefont{Battesti et~al.}(2004)\citenamefont{Battesti,
  Clad\'{e}, Guellati-Kh\`{e}lifa, Schwob, Gr\'{e}maud, Nez, Julien, and
  Biraben}}]{Battesti04}
\bibinfo{author}{\bibfnamefont{R.}~\bibnamefont{Battesti}},
  \bibinfo{author}{\bibfnamefont{P.}~\bibnamefont{Clad\'{e}}},
  \bibinfo{author}{\bibfnamefont{S.}~\bibnamefont{Guellati-Kh\`{e}lifa}},
  \bibinfo{author}{\bibfnamefont{C.}~\bibnamefont{Schwob}},
  \bibinfo{author}{\bibfnamefont{B.}~\bibnamefont{Gr\'{e}maud}},
  \bibinfo{author}{\bibfnamefont{F.}~\bibnamefont{Nez}},
  \bibinfo{author}{\bibfnamefont{L.}~\bibnamefont{Julien}}, \bibnamefont{and}
  \bibinfo{author}{\bibfnamefont{F.}~\bibnamefont{Biraben}},
  \bibinfo{journal}{Phys. Rev. Lett.} \textbf{\bibinfo{volume}{92}},
  \bibinfo{pages}{253001} (\bibinfo{year}{2004}).

\bibitem[{\citenamefont{Coq et~al.}(2005)\citenamefont{Coq, Retter, Richard,
  Aspect, and Bouyer}}]{Coq05}
\bibinfo{author}{\bibfnamefont{Y.~L.} \bibnamefont{Coq}},
  \bibinfo{author}{\bibfnamefont{J.~A.} \bibnamefont{Retter}},
  \bibinfo{author}{\bibfnamefont{S.}~\bibnamefont{Richard}},
  \bibinfo{author}{\bibfnamefont{A.}~\bibnamefont{Aspect}}, \bibnamefont{and}
  \bibinfo{author}{\bibfnamefont{P.}~\bibnamefont{Bouyer}},
  \bibinfo{journal}{ArXiv: cond-mat/0501520}  (\bibinfo{year}{2005}).

\bibitem[{\citenamefont{R.~S. Van~Dyck et~al.}(1987)\citenamefont{R.~S.
  Van~Dyck, Schwinberg, and Dehmelt}}]{VanDyck87}
\bibinfo{author}{\bibnamefont{R.~S. Van~Dyck}},
  \bibinfo{author}{\bibfnamefont{P.~B.} \bibnamefont{Schwinberg}},
  \bibnamefont{and} \bibinfo{author}{\bibfnamefont{H.~G.}
  \bibnamefont{Dehmelt}}, \bibinfo{journal}{Phys.\ Rev.\ Lett}
  \textbf{\bibinfo{volume}{59}}, \bibinfo{pages}{26} (\bibinfo{year}{1987}).

\bibitem[{\citenamefont{Kinoshita}(1995)}]{Kinoshita95}
\bibinfo{author}{\bibfnamefont{T.}~\bibnamefont{Kinoshita}},
  \bibinfo{journal}{Phys.\ Rev.\ Lett} \textbf{\bibinfo{volume}{75}},
  \bibinfo{pages}{4728} (\bibinfo{year}{1995}).

\bibitem[{\citenamefont{Hughes and Kinoshita}(1999)}]{Hughes99}
\bibinfo{author}{\bibfnamefont{V.~W.} \bibnamefont{Hughes}} \bibnamefont{and}
  \bibinfo{author}{\bibfnamefont{T.}~\bibnamefont{Kinoshita}},
  \bibinfo{journal}{Rev. Mod. Phys} \textbf{\bibinfo{volume}{71}},
  \bibinfo{pages}{S133} (\bibinfo{year}{1999}).

\bibitem[{\citenamefont{Schneble et~al.}(2003)\citenamefont{Schneble, Torii,
  Boyd, Streed, Pritchard, and Ketterle}}]{Schneble03}
\bibinfo{author}{\bibfnamefont{D.}~\bibnamefont{Schneble}},
  \bibinfo{author}{\bibfnamefont{Y.}~\bibnamefont{Torii}},
  \bibinfo{author}{\bibfnamefont{M.}~\bibnamefont{Boyd}},
  \bibinfo{author}{\bibfnamefont{E.~W.} \bibnamefont{Streed}},
  \bibinfo{author}{\bibfnamefont{D.~E.} \bibnamefont{Pritchard}},
  \bibnamefont{and} \bibinfo{author}{\bibfnamefont{W.}~\bibnamefont{Ketterle}},
  \bibinfo{journal}{Science} \textbf{\bibinfo{volume}{300}},
  \bibinfo{pages}{475} (\bibinfo{year}{2003}).

\bibitem[{\citenamefont{Castin and Dum}(1996)}]{Castin96}
\bibinfo{author}{\bibfnamefont{Y.}~\bibnamefont{Castin}} \bibnamefont{and}
  \bibinfo{author}{\bibfnamefont{R.}~\bibnamefont{Dum}},
  \bibinfo{journal}{Phys.\ Rev.\ Lett} \textbf{\bibinfo{volume}{77}},
  \bibinfo{pages}{5315} (\bibinfo{year}{1996}).

\bibitem[{\citenamefont{Meystre}(2001)}]{Meystre01}
\bibinfo{author}{\bibfnamefont{P.}~\bibnamefont{Meystre}},
  \emph{\bibinfo{title}{Atom Optics}} (\bibinfo{publisher}{Springer-Verlag},
  \bibinfo{address}{New York}, \bibinfo{year}{2001}).

\bibitem[{\citenamefont{Gupta et~al.}(2001)\citenamefont{Gupta, Leanhardt,
  Cronin, and Pritchard}}]{Gupta01}
\bibinfo{author}{\bibfnamefont{S.}~\bibnamefont{Gupta}},
  \bibinfo{author}{\bibfnamefont{A.~E.} \bibnamefont{Leanhardt}},
  \bibinfo{author}{\bibfnamefont{A.~D.} \bibnamefont{Cronin}},
  \bibnamefont{and} \bibinfo{author}{\bibfnamefont{D.~E.}
  \bibnamefont{Pritchard}}, \bibinfo{journal}{C.R. Acad. Sci. IV-Phys.}
  \textbf{\bibinfo{volume}{2}}, \bibinfo{pages}{479} (\bibinfo{year}{2001}).

\bibitem[{\citenamefont{Trippenbach et~al.}(2000)\citenamefont{Trippenbach,
  Band, Edwards, Doery, Julienne, Hagley, Deng, Kozuma, Helmerson, Rolston
  et~al.}}]{Trippenbach00}
\bibinfo{author}{\bibfnamefont{M.}~\bibnamefont{Trippenbach}},
  \bibinfo{author}{\bibfnamefont{Y.~B.} \bibnamefont{Band}},
  \bibinfo{author}{\bibfnamefont{M.}~\bibnamefont{Edwards}},
  \bibinfo{author}{\bibfnamefont{M.}~\bibnamefont{Doery}},
  \bibinfo{author}{\bibfnamefont{P.~S.} \bibnamefont{Julienne}},
  \bibinfo{author}{\bibfnamefont{E.~W.} \bibnamefont{Hagley}},
  \bibinfo{author}{\bibfnamefont{L.}~\bibnamefont{Deng}},
  \bibinfo{author}{\bibfnamefont{M.}~\bibnamefont{Kozuma}},
  \bibinfo{author}{\bibfnamefont{K.}~\bibnamefont{Helmerson}},
  \bibinfo{author}{\bibfnamefont{S.~L.} \bibnamefont{Rolston}},
  \bibnamefont{et~al.}, \bibinfo{journal}{J. Phys. B}
  \textbf{\bibinfo{volume}{33}}, \bibinfo{pages}{47} (\bibinfo{year}{2000}).

\bibitem[{\citenamefont{Morice et~al.}(1995)\citenamefont{Morice, Castin, and
  Dalibard}}]{Morice95}
\bibinfo{author}{\bibfnamefont{O.}~\bibnamefont{Morice}},
  \bibinfo{author}{\bibfnamefont{Y.}~\bibnamefont{Castin}}, \bibnamefont{and}
  \bibinfo{author}{\bibfnamefont{J.}~\bibnamefont{Dalibard}},
  \bibinfo{journal}{Phys. Rev. A.} \textbf{\bibinfo{volume}{51}},
  \bibinfo{pages}{3896} (\bibinfo{year}{1995}).

\bibitem[{\citenamefont{Stenger et~al.}(1999)\citenamefont{Stenger, Inouye,
  Chikkatur, Stamper-Kurn, Pritchard, and Ketterle}}]{Stenger99}
\bibinfo{author}{\bibfnamefont{J.}~\bibnamefont{Stenger}},
  \bibinfo{author}{\bibfnamefont{S.}~\bibnamefont{Inouye}},
  \bibinfo{author}{\bibfnamefont{A.~P.} \bibnamefont{Chikkatur}},
  \bibinfo{author}{\bibfnamefont{D.~M.} \bibnamefont{Stamper-Kurn}},
  \bibinfo{author}{\bibfnamefont{D.~E.} \bibnamefont{Pritchard}},
  \bibnamefont{and} \bibinfo{author}{\bibfnamefont{W.}~\bibnamefont{Ketterle}},
  \bibinfo{journal}{Phys.\ Rev.\ Lett} \textbf{\bibinfo{volume}{82}},
  \bibinfo{pages}{4569} (\bibinfo{year}{1999}).

\bibitem[{\citenamefont{Schirotzek}(2004)}]{Schirotzek04}
\bibinfo{author}{\bibfnamefont{A.}~\bibnamefont{Schirotzek}}, Diploma thesis,
  \bibinfo{school}{Universit\"{a}t Hamburg} (\bibinfo{year}{2004}).

\end{thebibliography}
\end{document}